\documentclass{article}
\usepackage{spconf,amsmath,graphicx}

\usepackage{hyperref}

\usepackage{enumitem}
\setlist{nosep, leftmargin=14pt}

\usepackage{mwe} % to get dummy images

\DeclareMathOperator*{\argmin}{arg\,min}

\title{Multi-modal deformable image registration using untrained neural networks}
\name{Quang Luong Nhat Nguyen \qquad Ruiming Cao \qquad Laura Waller}
\address{University of California, Berkeley, Berkeley, CA 94720}
\begin{document}
%\ninept
%
\maketitle
\begin{abstract}
Image registration techniques usually assume that the images to be registered are of a certain type (e.g. single- vs. multi-modal, 2D vs. 3D, rigid vs. deformable) and there lacks a general method that can work for data under all conditions. We propose a registration method that utilizes neural networks for image representation. Our method uses untrained networks with limited representation capacity as an implicit prior to guide for a good registration. Unlike previous approaches that are specialized for specific data types, our method handles both rigid and non-rigid, as well as single- and multi-modal registration, without requiring changes to the model or objective function. We have performed a comprehensive evaluation study using a variety of datasets and demonstrated promising performance. \footnote{Code: \url{https://github.com/quang-nguyenln/mdirunn}.}
% We present an untrained and unsupervised model that leverages the coarse-to-fine process to control the granularity of the output motion, allowing the model to handle both rigid and non-rigid registration flexibly. Its architecture of two sequential multi-layer perceptrons (MLPs) helps the model register all kinds of data, regardless of single or multi-modal. 
% Compared with the classic pair-wise image registration technique in Advanced Normalization Tools (ANTs), our method outperforms the baseline at all data levels in terms of relative distance for 2D images while yielding better dice coefficient for the evaluated 3D medical images.
\end{abstract}
\begin{keywords}
Image registration, multi-modal, deformable registration, implicit neural representation
\end{keywords}
\section{Introduction}
\label{sec:intro}
% what is image registration, its importance and applications
Image registration is a fundamental problem in image processing, which aims to find the spatial transformation between pairs of images. For example, one might seek to register two images in which the object moved between frames~\cite{maintz1998survey}. In various fields in medical imaging and computer vision, image registration is commonly used to enable the integration and analysis of data across different sources, such as combining information between MRI and CT scans or environmental monitoring in augmented reality applications~\cite{zitova2003image}.

% single vs multi-modal reg
Single-modal registration - in which both images have the same contrast mechanism - has been extensively studied. Intensity-based similarity measures, such as mean square difference (MSD)~\cite{klein2009elastix}, have been developed to quantify the alignment between two images. In contrast, images from different modalities often involve complex and often non-linear pixel intensity correlation, whose alignment is thus much more difficult to quantify. As a result, there are fewer robust algorithms for multi-modal registration. Existing multi-modal registration algorithms face several drawbacks: mutual information (MI) suffers from low accuracy due to the overlooking of local structural information~\cite{bashiri2018multi}, and feature-based approaches often fail when corresponding structures have diverse appearances in different modalities~\cite{maes1997multimodality}.

% different types of motion in the registration
Another crucial distinction in image registration lies in the fundamental difference between rigid and deformable motions. Rigid registration encompasses global transformations such as translation, rotation, and scaling, and is typically solved by minimizing a similarity metric. Deformable registration, however, allows for local, non-linear deformations and is significantly more challenging. It requires estimating a dense displacement field that can vary at each spatial location, resulting in a much higher-dimensional optimization problem. Deformable methods often employ regularizers to ensure smoothness and physical plausibility. Popular approaches include free-form deformations using B-splines, diffeomorphic transformations, and, more recently, deep learning-based methods that directly predict displacement fields~\cite{balakrishnan2019voxelmorph}.

Previous research has divided image registration into multiple sub-categories, each addressed by highly specialized methods. Hence, end users must correctly determine which sub-category their dataset fits into. To address this limitation, there is a growing need for a more versatile and robust approach to image registration—one that can effectively handle various types of registrations.

% MODIFIED
\begin{figure*}[]
  \centering
\centerline{\includegraphics[width=\linewidth]{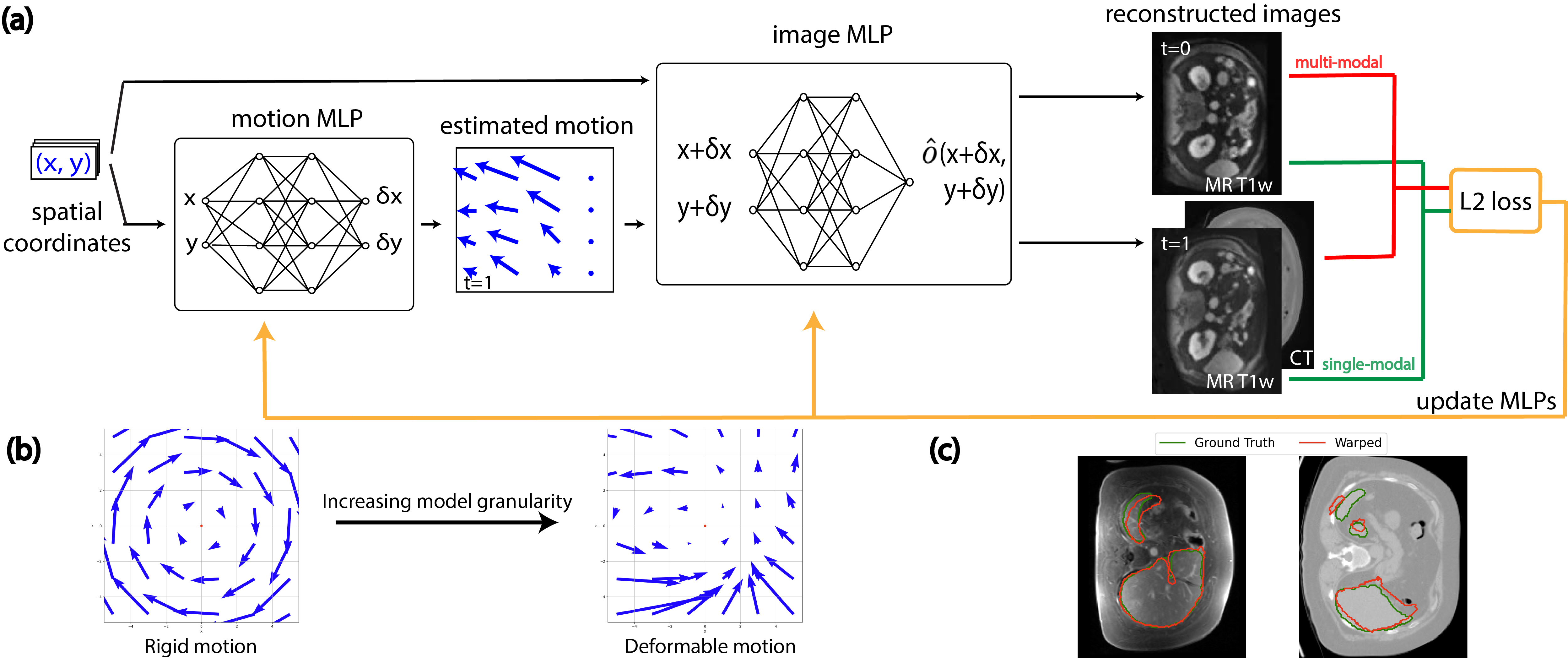}}
\caption{Untrained neural network for image registration. (a) The model outputs two channels, each representing a different modality, and two time points, denoting fixed and transformed input images. The L2 loss between the reconstructed images and their corresponding ground truth images is then computed and used to update the networks' weights. For single-modal registration, the L2 loss is applied on one output channel while in the multi-modal case, the loss function is computed using different output channels corresponding to the image modalities. The hash embedding is always applied to the input coordinate, so we consider it part of the network and omit it for visibility. (b) The model can handle rigid and deformable registration by tuning the granularity value. (c) An example of the 3D multi-modal registration of MRI-CT data. The green contours are the liver and spleen segmentations annotated on the displayed image, and the red contour refers to the segmentation annotated on the other image modality and warped using our method.}
\label{fig:model}
\end{figure*}
% learning based methods (data driven methods) for reg. non data driven, neural network methods (implicit neural represetnation)
Recently, neural networks trained on datasets with paired images and registration labels have demonstrated success in handling large deformations~\cite{balakrishnan2019voxelmorph}. Nonetheless, curating a large dataset with ground truth labels is often infeasible, particularly for multi-modal images. In the absence of such training dataset, an image-to-image network have been used to convert an image from one modality to the other, simplifying the problem to single-modal registration~\cite{tanner2018generative, lu2022image}. Alternatively, untrained coordinate-based neural networks are shown to be a good building block for image registration methods, providing a flexible way to parameterize the spatial transformation between two images~\cite{byra2023exploring}. Byra et al. \cite{byra2023exploring} has used such network to register single-modal 2D brain images. In our work, we extend to multi-modal deformable registration by combining two coordinate-based networks.

In this paper, we propose using untrained coordinate-based neural networks to register all kinds of data, regardless of single- vs multi-modal, rigid vs deformable assumption, and 2D vs 3D. We represent each pair of input images using two lightweight neural networks: one network storing the displacement map and the other storing the images. Due to the limited model capacity of these networks, a good registration must be optimized to efficiently store the input images into the network weights. Our method is general and versatile, using the same model and L2 loss function for all types of paired images, without requiring any prior or pre-training data. We validate our method with several experiments, including 2D single-modal data with rigid motion, 2D multi-modal data with rigid motion, and 3D multi-modal data with deformable motion.

\section{Materials and Methods}
\label{sec:materialsmethods}

\subsection{Coordinate-based Neural Networks}
\label{sec:coordnet}
A coordinate-based neural network, also known as an implicit neural representation, is a multi-layer perceptron that represents a multi-dimensional signal, such as a 2D image~\cite{sitzmann2020implicit} or a 3D volume~\cite{mildenhall2021nerf}, through its network weights. In the 2D case, a trained coordinate-based network takes in an arbitrary spatial coordinate $(x,y)$ and outputs the corresponding pixel value. Given an image, $I$, the network, $f\left( \cdot ; \theta \right)$, is trained to minimize the mean-square error between its output and the image pixel values for arbitrary spatial coordinates, $\left( x,y \right)$, such that 
\begin{equation}
\argmin_{\theta} \sum_{(x,y)} |f(x,y;\theta) - I(x,y)|^2.
\end{equation}
% Once the weight is optimized, any interested, on- and off-grid pixel can be obtained without additional interpolation or rounding by querying the corresponding coordinate from the MLP. 

\subsection{Image registration using untrained neural networks}
Our proposed model connects the reference and transformed images using two untrained coordinate-based networks: the motion and the image network. As shown in Fig.~\ref{fig:model}, the motion network ($f_{\text{m}}$) takes in a spatial coordinate $(x,y)$ and estimates the relative displacement $(\delta x, \delta y)$. By repeating this process for all pixels, we obtain a dense motion kernel, which contains the displacement for every coordinate. Because of the non-linearity from the activation functions, the motion network can represent global and locally deformable motion. When the fixed image is stored in the image network as described in Sec.~\ref{sec:coordnet}, we can obtain an image with arbitrary motion or distortion by querying the image network with spatial coordinates adjusted by various motion kernel. 

To produce a pixel value at $\left(x, y\right)$ for the transformed image, we add its spatial coordinate $\left( x, y \right)$ to the motion displacement output from the motion network before feeding it into the image network. Mathematically, the transformed image can be written as
$$I_\text{trans} = f_{im}\left( \left(x , y \right) + f_{mo} \left(x, y; \theta_{mo} \right); \theta_s \right),$$ where $f_{mo}$, $f_{im}$ is the motion and image networks. 

% To reconstruct the fixed image, we can directly query our desired spatial coordinates from the image network. 
% The new motion-transformed spatial coordinate $(x +\delta x, y + \delta y)$ is then mapped to another hash-embedded feature vector $\mathbf{h}(x + \delta x, y + \delta y)$ before being passed into the image network ($f_{\text{s}}$), producing the pixel value at location $(x, y)$ and time $t=1$. For $t=0$, we pass the hash-embedded feature of a spatial coordinate $\mathbf{h}(x,y)$ into the image network, obtaining the pixel value at $(x,y)$ for the fixed reconstructed image. All spatial coordinates undergo these two processes, allowing us to reconstruct a complete image at each time $t$. 

The weights of both networks are unknown a priori. For each new pair-wise registration, we randomly initialize the weights of two networks and optimize them to represent the fixed and transformed images using a L2 loss. Mathematically,
\begin{equation}
\begin{split}
\argmin_{\theta_{im}, \theta_{mo}} \sum_{\left( x, y\right)} & |f_{im} \left( x , y; \theta_{im} \right) - I_\text{ref}^{\left( x, y\right)} |^2 + \\
& | f_{im} \left( \left(x , y \right) + f_{mo} \left(x, y; \theta_{mo} \right); \theta_{im} \right) - I_\text{trans}^{\left( x, y\right)} |^2 .
\end{split}
\end{equation}
The image network outputs one channel for single-modal and two channels for multi-modal registration. For the multi-modal case, the first and second terms of the loss function are applied on different output channels, one for each image modality.

\subsection{Model Granularity}
To increase network representation capacity, we apply a hash encoding method to map the input spatial coordinate into multi-resolution hash-embedded features, $\mathbf{h} = [h_0, \cdots, h_{N-1}]$, before feeding them into the coordinate-based networks~\cite{muller2022instant}. A coarse resolution feature (for example, $h_0$) will vary much slower than a finer resolution feature (for example, $h_{N-1}$) when the input coordinate changes. Thus, having only coarse resolution features as the input can limit the model's capacity to represent complex signal.

We leverage this multi-resolution granularity control to tune the model capacity for both networks. We adjust the fine resolution of the embedding for the motion network to accommodate for rigid vs deformable registration. As illustrated in Fig.~\ref{fig:model}b, for rigid motion, we set the finest resolution to be the same as the coarsest resolution, so the the motion kernel output is less spatially varying and more rigid, and vice versa. 

In multi-modal registration, the L2 loss cannot sufficiently guide to a good registration by itself since the two images are in separate channels. We use a coarse-to-fine strategy that re-weighs the embedded features in the optimization process. We set the capacity of both networks to be very restricted at the start of training. To efficiently store the two images into the image network with limited capacity, the motion network has to better align them for a maximized correlation of pixel values in two modalities. The capacity is gradually increased as the optimization converges. In our implementation, a feature, $h_i$, is re-weighted by $\text{clip}\left(N\alpha-i, 0, 1\right)$. Here, $\alpha = \text{min} \left( 1, \frac{e}{e_g} \right)$ sets the current coarse-to-fine level, where $e$ is the current epoch of the reconstruction and $e_g$ is the target epoch to finish the coarse-to-fine process. 
% of the hash embedding is controlled by the motion granularity, , . With a nonzero value of $e_g$, % This forces the motion MLP to learn a sufficient deformation field such that the image network can efficiently store the input images with its weights. In other words, the proposed model can effectively handle rigid and deformable (non-rigid) motion by tuning $e_g$.
% \subsection{Motion Skipping}
% If the space-time coordinates $(x,y,t=0)$ are also fed into the motion and image network sequentially to reconstruct the fixed image, the model may converge to a local minima as the image network can overfit to the image before the motion is fully recovered. Given their interdependence, the networks' failure to align their updates will result in pixel-perfect reconstructed images but arbitrarily scaled motion. To mitigate this issue, the spatial coordinates $(x,y)$ is passed directly into the image MLP without going through the motion MLP for the fixed image reconstruction. 
% \subsection{Single-Modal vs. Multi-Modal}

\section{Results}
\label{sec:results}
\subsection{Datasets and metrics}
We evaluate our method using two datasets. The Zurich dataset~\cite{lu2022image} contains 108 QuickBird-acquired 2D satellite image patches of the city of Zurich. Each image patch is extracted as two modalities: near-infrared (NIR) channel and the joint RGB channel. This dataset contains four levels of rigid motion with varying motion magnitude, where a higher level implies greater motion~\cite{lu2022image}. Single-modal registration uses images in NIR channel, and multi-modal registration takes NIR channel for the fixed image and joint RGB channel for the transformed one. 
We follow the evaluation protocol in~\cite{lu2022image} for both single-modal and multi-modal registration on the Zurich dataset.: we first measure the distance between the spatial coordinates obtained after the registration and ground truth coordinates for each corner of the transformed image. Then, we take the average distance over four corners and divide it by the image height/width for \textit{relative distance}. We consider a registration successful if $\Delta < 2$ and compute the success rate as the ratio of success cases to total cases.

For deformable registration, we used the Abdomen MR-CT dataset~\cite{clark2013cancer, kavur2021chaos}, which consists of 16 pairs of MR-CT scans with manual segmentations of different organs and arbitrary deformable motion between the images. We use the eight pairs of MR-CT scans from the Abdomen MR-CT dataset that have ground truth labels for our evaluation. For quantitative evaluation, we use dice similarity coefficient (DSC) which measures the volumetric overlap between the ground truth segmentation and the estimated segmentation, and HD-95 which assesses $95^{\text{th}}$ percentile of the distances between the points of the ground truth and the predicted labels. 

\subsection{Single-Modal Registration}
\vspace{-0.4cm}
\begin{table}[h]
\setlength{\tabcolsep}{3pt}
\caption{Evaluation for single-modal and rigid registration of the proposed method vs. baseline method (Rigid + L2) on 2 metrics: Success rate (0-1) and Relative Distance (in \%).}
\footnotesize
\begin{tabular}{lcccccccc}
\hline
 & \multicolumn{2}{c}{Level 1}  & \multicolumn{2}{c}{Level 2} & \multicolumn{2}{c}{Level 3} & \multicolumn{2}{c}{Level 4}   \\
  & \multicolumn{1}{c}{$Suc.$} & \multicolumn{1}{c}{$Rel.$} & \multicolumn{1}{c}{$Suc.$} & \multicolumn{1}{c}{$Rel.$} & \multicolumn{1}{c}{$Suc.$} & \multicolumn{1}{c}{$Rel.$} & \multicolumn{1}{c}{$Suc.$} & \multicolumn{1}{c}{$Rel.$} \\
  & \multicolumn{1}{c}{$Rate$} & \multicolumn{1}{c}{$Dist.$} & \multicolumn{1}{c}{$Rate$} & \multicolumn{1}{c}{$Dist.$} & \multicolumn{1}{c}{$Rate$} & \multicolumn{1}{c}{$Dist.$} & \multicolumn{1}{c}{$Rate$} & \multicolumn{1}{c}{$Dist.$}\\
\hline
& $0.44$ & $2.33$ & $0.38$ & 6.30 & $0.13$ & $12.9$ & $0.01$ & 19.9 \\[-2ex]
\raisebox{2ex}{Rigid + L2}  &  & (1.53) & & (4.79) & & (7.05) & & (6.54)\\[0ex]
& $\mathbf{1.00}$ & $0.01$ & $\mathbf{0.79}$ & 1.08 & $\mathbf{0.22}$ & $7.62$ & $\mathbf{0.05}$ & 15.25  \\[-2ex]
\raisebox{2ex}{Our method} & & (0.01) & & (2.20) & & (4.59) & & (4.21)\\[0ex]
\hline
\end{tabular} 
% \end{center}
\label{smr}
\vspace{-0.3cm}
\end{table}

\begin{figure}[htb]
  \centering
\centerline{\includegraphics[width=0.7\linewidth]{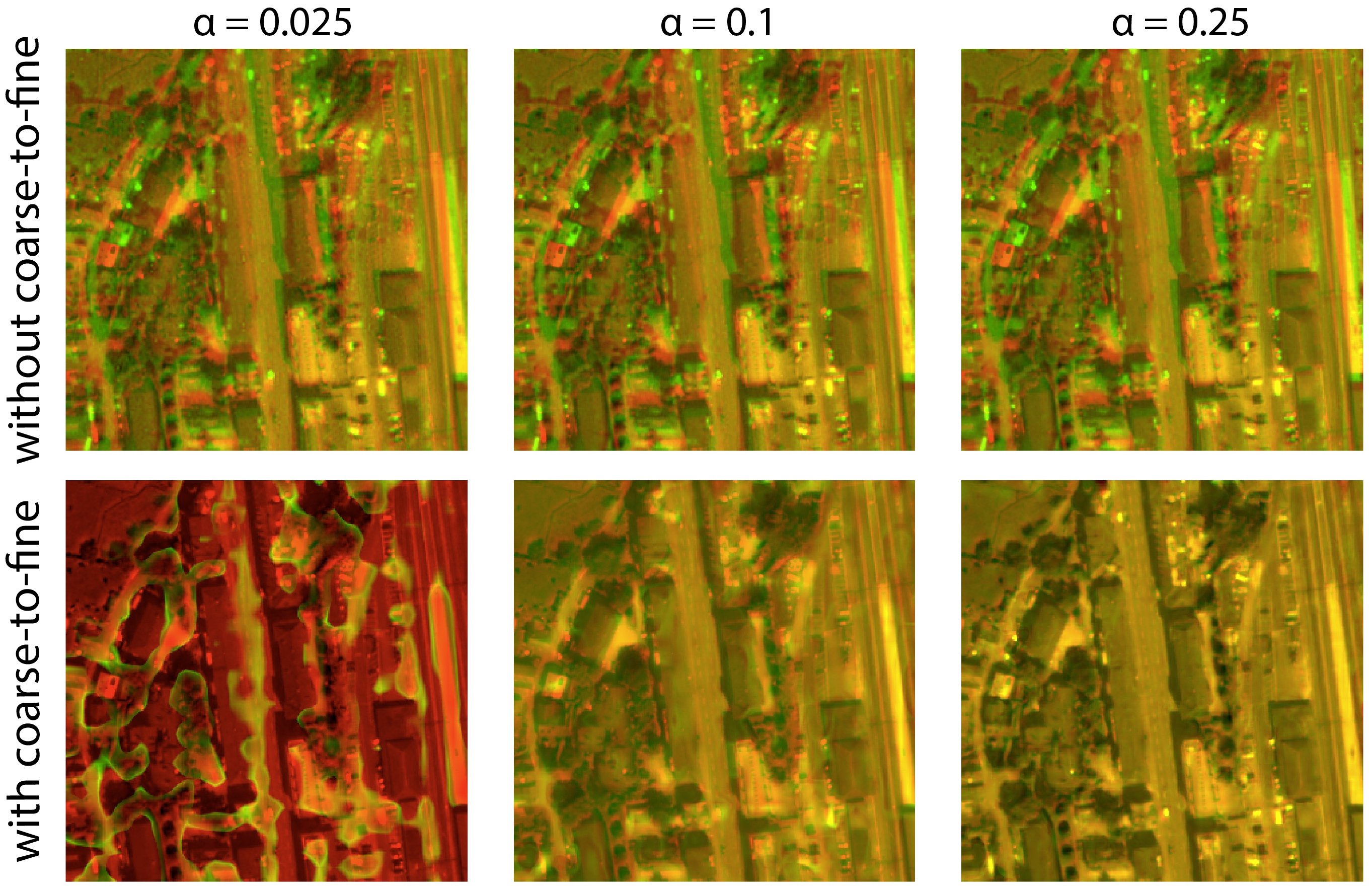}}
%  \vspace{2.0cm}
\caption{The network reconstruction of fixed image after applying motion kernel (green channel) overlaid with the transformed image (red) in the multi-modal case. The top row does not use the coarse-to-fine process, and the two images are not properly registered. The bottom row with coarse-to-fine produces coarser reconstruction at the beginning but eventually aligns multi-modal images.}
\label{fig:ctf}
% \vspace{-0.4cm}
\end{figure}

The quantitative results for single-modal registration are presented in Table~\ref{smr}. We compare our method against rigid registration method with meansquares as the optimization metric from the Advanced Normalization Tools (ANTs) library~\cite{avants2011reproducible}. Our method successfully registers all pairs of image patches of level 1 and has 79\% and 38\% success rates for levels 2 and 3, 41\% and 9\% higher than the baseline method. Across all transformation levels, the model achieves a significantly higher success rate and performs better than the baseline method for average relative distance. 
\subsection{Multi-Modal Registration}
% \vspace{-0.4cm}
\begin{table}[h]
\setlength{\tabcolsep}{3pt}
{
\caption{Evaluation for 2D multi-modal and rigid registration of the proposed method vs. baseline method (Rigid + MI) on two metrics: Success rate (0-1) and Relative Distance (in \%)}
\footnotesize
\begin{tabular}{lcccccccc}
\hline
 & \multicolumn{2}{c}{Level 1}  & \multicolumn{2}{c}{Level 2} & \multicolumn{2}{c}{Level 3} & \multicolumn{2}{c}{Level 4}   \\
  & \multicolumn{1}{c}{$Suc.$} & \multicolumn{1}{c}{$Rel.$} & \multicolumn{1}{c}{$Suc.$} & \multicolumn{1}{c}{$Rel.$} & \multicolumn{1}{c}{$Suc.$} & \multicolumn{1}{c}{$Rel.$} & \multicolumn{1}{c}{$Suc.$} & \multicolumn{1}{c}{$Rel.$} \\
  & \multicolumn{1}{c}{$Rate$} & \multicolumn{1}{c}{$Dist.$} & \multicolumn{1}{c}{$Rate$} & \multicolumn{1}{c}{$Dist.$} & \multicolumn{1}{c}{$Rate$} & \multicolumn{1}{c}{$Dist.$} & \multicolumn{1}{c}{$Rate$} & \multicolumn{1}{c}{$Dist.$}\\
\hline
& $0.36$ & $3.56$ & $0.14$ & 9.29 & $0.01$ & $16.04$ & $0$ & 22.96 \\[-2ex]
\raisebox{2ex}{Rigid + MI} &  & (3.47) & & (4.94) & & (4.48) & & (3.92)\\[0ex]
& $\mathbf{0.82}$ & $1.16$ & $\mathbf{0.35}$ & 5.72 & $\mathbf{0.11}$ & $11.51$ & $0$ & 18.25   \\[-2ex]
\raisebox{2ex}{Our method} & & (3.35) & & (4.76) & & (4.44) & & (2.28) \\[0ex]
\hline
\end{tabular} }
\label{mmr}
% \vspace{-0.3cm}
\end{table}
The quantitative evaluation of 2D multi-modal and rigid registration is shown in Table~\ref{mmr}. We use the rigid registration setting of ANTs with MI as the optimization metric~\cite{avants2011reproducible} for the baseline comparison. The proposed method achieved significantly higher success rates of 82\% and 35\% for levels 1 and 2, 46\% and 21\% higher than the baseline. While the baseline method fails to register any image patches with more significant transformations, our network succeeds in 11\% of the cases for level 3. The model's low success rate for the higher levels is because of the drastic mismatching of features between the fixed and transformed image patches, small image size ($300\times 300$), and the stark contrast in intensities between the two modalities.

\begin{table}[h]
\centering
{\caption{Evaluation results for 3D multi-modal and deformable registration of the proposed method vs. baseline methods. Numbers are reported as avg$\pm$std.}
\footnotesize
\begin{tabular}{lccc}
\hline
 & SyN (MI) & SyN (CC) & Our method \\
\hline
Average DSC & $0.41 \pm 0.21$ & $0.44 \pm 0.23$ & $\mathbf{0.49 \pm 0.17}$ \\
Weighted DSC & $0.51 \pm 0.15$ & $0.52 \pm 0.14$ & $\mathbf{0.54 \pm 0.14}$\\
HD-95 (mm) & $19.6 \pm 11.1$ & $\mathbf{18.5 \pm 11.2}$ & $18.9 \pm 10.1$\\
\hline
\end{tabular} }
\label{mmd}
\end{table}

% MODIFIED
\begin{figure}[htb]
  \centering
\includegraphics[width=0.6\linewidth]{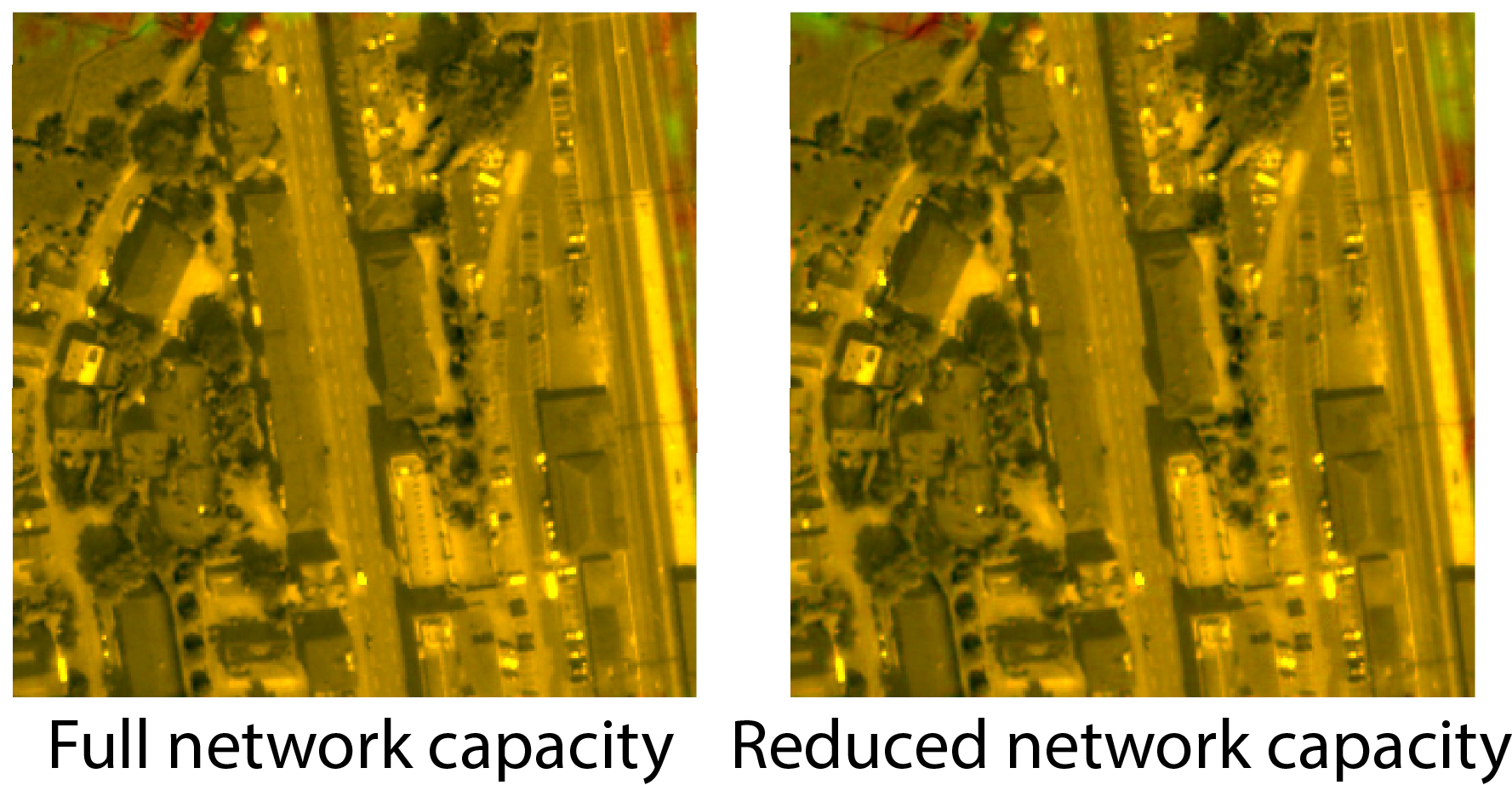}
%  \vspace{2.0cm}
\caption{The network reconstruction of fixed image after applying motion kernel (green channel) overlaid with the transformed image (red). The left image is registered using an image network with full representation capacity, while the right image is registered with an image MLP with reduced capacity.}
\label{fig:res}
\end{figure}
For 3D multi-modal and deformable registration, we use symmetric normalization SyN (affine and deformable transformation) with mutual information (MI) and cross-correlation (CC) as the optimization metric as the baseline methods~\cite{avants2008symmetric}. As shown in Table \ref{mmd}, our proposed method achieves a higher average and weighted Dice than both settings of ANTs. However, it receives a slightly worse HD-95, $0.4$ mm higher, than the baseline method that uses CC.

Our model is able to register multi-modal images without explicitly optimizing for the pixel correlation. As illustrated in Fig.~\ref{fig:ctf}, without restricting the network capacity in the earlier training, the image MLP is prone to overfit to both input images too quickly before the motion MLP can account for any motion. However, with the coarse-to-fine process, the model capacity is limited in the beginning, forcing the motion network to search for the best alignment and resulting in a good registration.

% \vspace{-0.4cm}
\section{Discussion}
\label{sec:discussion}
% \vspace{-0.2cm}
% MODIFIED
We have presented a general image registration method which uses untrained coordinate-based network as an implicit prior to guide for good alignment. With a simple L2 loss function, our method allows for robust registration of different kinds of data, be it single- or multi-modal, rigid or deformable motion, and 2D or 3D. In our evaluation using multiple data types, our methods outperformed conventional registration methods specialized on each of these data types. 

We simulate the extreme situation where the image contains more information than what the image network can store by limiting its granularity to $\frac{1}{15}$ the original resolution. Fig.~\ref{fig:res} shows that our method maintains robust registration performance in such situation.

% MODIFIED
Our findings are subjected to several challenges. Since the implicit networks are optimized from scratch for each new image pair, their runtime is longer than convolutional networks trained with paired data. It would be interesting to investigate how hyper-networks can be utilized to pre-initialize the weights of the networks or how MLPs can be pre-trained on population data~\cite{ha2016hypernetworks}.  Furthermore, incorporating more sophisticated regularizers for weights optimization could prevent convergence to non-smooth and non-invertible deformation fields~\cite{zhang2013bayesian} and enable the model to better capture larger and more complex deformations.

% Below is an example of how to insert images. Delete the ``\vspace'' line,
% uncomment the preceding line ``\centerline...'' and replace ``imageX.ps''
% with a suitable PostScript file name.
% -------------------------------------------------------------------------
% \begin{figure*}[htb]
% \centering
%     \begin{minipage}{\textwidth}
%     \centering
% \begin{minipage}[c]{0.3\linewidth}
%   \centering
%   \centerline{\includegraphics[width=8.5cm]{example-image}}
% %  \vspace{2.0cm}
%   \centerline{(a) Result 1}\medskip
% \end{minipage}
% %
% \hfill
% \begin{minipage}[c]{.3\linewidth}
%   \centering
%   \centerline{\includegraphics[width=4.0cm]{example-image}}
% %  \vspace
%   \centerline{(b) Results 3}\medskip
%   \vspace{0.5cm}
%   \centering
%   \centerline{\includegraphics[width=4.0cm]{example-image}}
% %  \vspace{1.5cm}
%   \centerline{(c) Result 4}\medskip
% \end{minipage}
% \begin{minipage}[c]{.3\linewidth}
%   \centering
%   \centerline{\includegraphics[width=4.0cm]{example-image}}
% %  \vspace
%   \centerline{(b) Results 3}\medskip
%   \vspace{0.5cm}
%   \centering
%   \centerline{\includegraphics[width=4.0cm]{example-image}}
% %  \vspace{1.5cm}
%   \centerline{(c) Result 4}\medskip
% \end{minipage}
% \end{minipage}
% \caption{Example of placing a figure with experimental results.}
% \label{fig:res}
% %
% \end{figure*}

% \vfill
% \pagebreak

\section{Acknowledgments}
\label{sec:acknowledgments}
This work has been made possible in part by CZI grant DAF2021-225666 and grant DOI \href{https://doi.org/10.37921/192752jrgbnh}{10.37921/192752jrgbnh} from the Chan Zuckerberg Initiative DAF, an advised fund of Silicon Valley Community Foundation (funder DOI \href{https://doi.org/10.13039/100014989}{10.13039/ 100014989}). R.C. was supported in part by Siebel Scholarship. L.W. is a Chan Zuckerberg Biohub SF Investigator.

% IEEE-ISBI supports the disclosure of financial support for the project
% as well as any financial and personal relationships of the author that
% could create even the appearance of bias in the published work. The
% authors must disclose any agency or individual that provided financial
% support for the work as well as any personal or financial or
% employment relationship between any author and the sources of
% financial support for the work.

% Other types of acknowledgements can also be listed in this section.

% Reporting on real or potential conflicts of interests, or the absence
% thereof, is required in the paper. Authors are responsible for
% correctness of the statements provided in the manuscript. Examples of
% appropriate statements include:
% \begin{itemize}
%   \item ``No funding was received for conducting this study. The
%     authors have no relevant financial or non-financial interests to
%     disclose.'' 
%   \item ``This work was supported by […] (Grant numbers) and
%     […]. Author X has served on advisory boards for Company Y.'' 
%   \item ``Author X is partially funded by Y. Author Z is a Founder and
%     Director for Company C.''
% \end{itemize}

\bibliographystyle{IEEEbib}
\bibliography{strings,refs}

\end{document}